\documentclass[eprint, twocolumn]{revtex4-2}
\usepackage{bbm}
\usepackage{natbib}
\usepackage{graphics}
\usepackage{amsmath}
\usepackage{mathrsfs}
\usepackage{theorem}
\usepackage{float}
\usepackage{hyperref}
\usepackage{rotating}
\usepackage{amssymb}
\usepackage[english]{babel}
\usepackage{color}
\usepackage{fancybox}
\newcommand{\ket}[1]{|#1\rangle}

\newcommand{\be}{\begin{eqnarray}}
\newcommand{\ee}{\end{eqnarray}}

\begin{document}

\title{Anomalous mobility edges and extended-localized transition in a quasiperiodic  emitter-cavity array}
\author{H. T. Cui $^{1}$}
\email{cuiht01335@aliyun.com}
\author{H. Z. Shen $^{2}$}
\email{shenhz458@nenu.edu.cn}
\author{M. Qin $^{1}$}
\email{qinming@ldu.edu.cn}
\author{X. X. Yi $^{2}$}
\email{yixx@nenu.edu.cn}
\affiliation{$^1$ School of Physics and Optoelectronic Engineering \& Institute of Theoretical Physics, Ludong University, Yantai 264025, China}
\affiliation{$^2$ Center for  Quantum Sciences, Northeast Normal University, Changchun 130024, China}
\date{\today}

\begin{abstract}
The manipulation of localization in quasiperiodic systems by mobility edges or localization transition holds significant physical importance. In this letter, we demonstrated  that the dissipation  can induce the  emergence  of anomalous mobility edges  and  extended-localized transition in emitter-cavity arrays  controlled by quasiperiodic potentials.  Specifically, we observe  that the localization properties of emitters is governed by the nature of quantum bound states, either discrete or embedded in continuum, providing a unified mechanism linking the emitter-photon bound physics to quasiperiodic criticality. Depending on the bound state discrete or continuumlike,  the induced effective excitation hopping exhibits either exponentially decaying or sinusoidally oscillating, giving rise to  the formation of localized or critical states, respectively. Through a generalized duality transformation, we analytically determine the anomalous mobility edges  and the critical strength of potential, enabling the construction of a full phase diagram. The study reveals that  the physical characteristics of cavity exert a significant  influence on excitation localization. Therefore, the manipulation of excitation localization can be achieved solely by adjusting the cavity fields. 
\end{abstract}

\maketitle 

Quasiperiodic quantum systems are of broad interest in condensed-matter physics and quantum information processing due to their exotic transport properties \cite{abanin}.  Unlike disordered systems, they exhibit a number of unique and experimentally relevant features. In one-dimensional quasiperiodic systems, the extended-localized  transition can occur \cite{aa, haper, basko, huse}, whereas in random disordered systems, this transition is limited to three dimensions.  Furthermore, one-dimensional quasiperiodic systems  can feature a mobility edge (ME) separating localized and delocalized states, which can be precisely determined through duality transformation\cite{sarma90,biddle,ganeshan}.  A crucial advancement includes the introduction of the global theory to analytically determine MEs when duality is breaking \cite{wang20}.  Recently it have revealed that one-dimensional quasiperiodic systems can display a critical phase, which behaves neither localized nor extended \cite{wang16, wang21, wang22, liu22, zhou23, goncalves23, li23, banerjee, zhou26, huang26}.   The usage of global theory has led to the exact determination of an anomalous ME, separating critical states from localized and extended states \cite{liu22, zhou23, banerjee, zhou26}. Experimental verification of the critical phase and anomalous MEs has been achieved in a multiple superconducting qubit quantum system \cite{huang26} and  in a quasi-periodically driven orbital optical lattice with ultracold atoms \cite{hu26}.  

In the presence of dissipation, localization is generally expected to become unstable.   Surprisingly, recent research has shown that dissipation can lead to an extended-localized transition in the one-dimensional mosaic model \cite{liu24}, challenging the conventional view that dissipation merely disrupts localization. This study highlights that tailored jump operators can maintain MEs and thus protect localization feature  against dissipation, significantly impacting the system's stationary properties. In contrast,  another recent study has shown that strong dephasing can trigger the emergence of MEs in a quasiperiodic system, that typically does not exhibit such feature \cite{longhi}. Conversely, in systems initially possessing MEs, the dephasing effect destroys MEs, and thereby degrades the localization feature of system.  These  results suggest that the localization behavior of quasiperiodic quantum systems can be actively controlled by appropriately engineering their environment.

The above studies raise two key questions. First, the mechanism behind decoherence-induced localization remains unclear.  Ref. \cite{liu24} suggests  that the emergence of an extended–localized transition is closely tied to the form of the jump operator: MEs are preserved only when the jump operator displays the same spatial structure as the onsite potential.  Similarly, Ref. \cite{longhi} demonstrates that dephasing-induced MEs  manifest only when the system initially lacks intrinsic MEs and is subjected  to sufficiently strong dephasing. These observations suggest that the decoherence-induced localization  is not universal, but instead emerges from a subtle interplay between the quasiperiodic potential and the type or strength of decoherence. Nevertheless, a precise microscopic mechanism remains to be established. Secondly, the inquiry into whether decoherence can induce critical states and anomalous MEs is of practical importance, particularly in light of experimental evidence of critical states\cite{huang26, hu26}. Given that no quantum system is immune to decoherence, identifying the conditions for the emergence of decoherence-induced critical states and anomalous MEs holds significant physical interest.      

\begin{figure}
\center
\includegraphics[width=8cm]{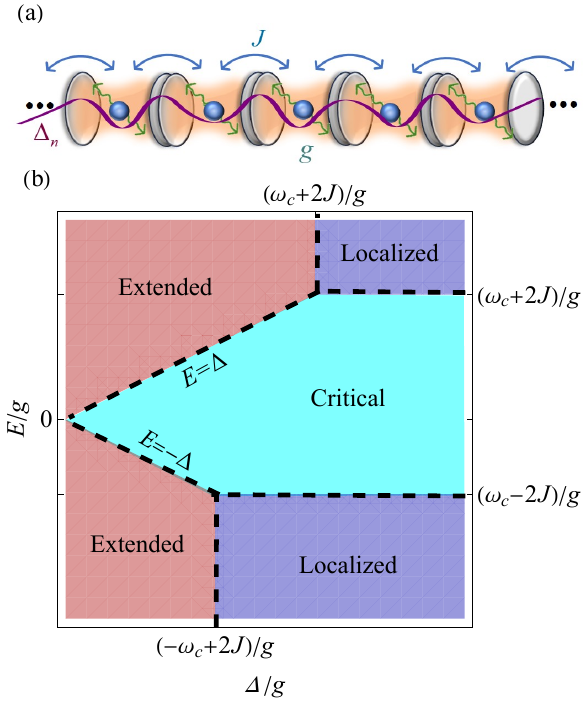}
\caption{(Color online) (a) Schematic for the emitter-cavity chain described by Eq. \eqref{H}. Blue spheres denote the emitters. The purple wave line represents the quasiperiodic potentials imposed on emitters. The blue curves with two arrows depicts the hopping of photon. Green wave line  denotes the coupling between emitter and cavity. (b) The phase diagram determined by evaluating the localization feature of the effective energy levels for the emitters. The dashed-black lines mark the boundaries between extended (orange), localized (blue) and critical (cyan) phases. The physical meaning for all symbols can be found in the maintext. }
\label{fig:phase}
\end{figure}

In this study, we address these questions by investigating  the localization properties of single excitation in an emitter-cavity array  subjected to quasiperiodic potentials \cite{greentree06, angelakis07}, as depicted in Fig.\ref{fig:phase}(a). By analyzing  the emitter-photon bound states,  we demonstrate that the long-time single excitation dynamics   can exhibit  delocalized, localized, or critical characteristics, determined by the nature of bound states,  discrete or embedded in the continuum of cavities. The underlying physics  is that  the emergence of discrete or continuum-like bound state can induce the effective excitation hopping exhibiting either exponentially decaying or sinusoidally oscillating, which give rise to the distinct localization properties of excitation. Owing to this effective hopping, the anomalous MEs and delocalized-localized transition can be identified analytically via  a duality transformation.  Based on this analysis, we construct the full phase diagram illustrated in Fig. \ref{fig:phase} (b).

\emph{The Model} We consider one-dimensional $N$ two-level emitters which experience a quasiperiodic onsite potential, interacting individually to an array of coupled cavities with single mode, as depicted  by  Fig.\ref{fig:phase}(a).  The total Hamiltonian  can be written as ($\hbar=1$) 
\be \label{H}
H=H_s + H_c + H_{\text{int}},
\ee
in which, 
\be 
H_s&=& \sum_{n=1}^{N} \Delta_n \sigma^{+}_n \sigma^{-}_n, \nonumber \\
H_b&=&  \omega_c\sum_{j=1}^{N} b_j^{\dagger}b_j + J \sum_{j=1}^{N-1} \left(b_j^{\dagger} b_{j+1} +b_{j+1}^{\dagger} b_j\right), \\
 H_{\text{int}} &=& g \sum_{n, j=1}^{N} \left(\sigma^{+}_n b_{j} +b_{j}^{\dagger} \sigma^{-}_n\right)\delta_{n,j}, \nonumber
\ee
where $\delta_{n,j}$ denotes the Kronecker $\delta$-function. $H_s$ depicts the emitters manipulated by the quasiperiodic onsite potential $\Delta_n=\Delta \cos\left(2\pi \beta n\right)$ with the Golden ratio $\beta=\left(\sqrt{5}-1\right)/2$. $\sigma_n^{+\left(- \right)}$ represents the raising (lowering) operator for the $n$-th emitter. The cavities  are coupled with strength $J$  to enable  photon transport along the cavity array. $J\equiv 1 $ is supposed by default in the following discussion. The photonic dynamics is   captured by a tight-binding Hamiltonian  $H_b$, where $\omega_c$ is the center frequency of cavity. $b_j^{\left( \dagger \right)}$ denotes the bosonic annihilation (creation) operator for the $j$-th cavity.  $H_b$ exhibits a continuum band of width $4J$ centered at $\omega_c$, with the dispersion relation $\omega_k=\omega_c + 2J \cos k \left( k\in\left[-\pi, \pi\right]\right)$.  The emitter indexed as  $n$ is locally coupled  to the corresponding cavity $n$ through $H_{\text{int}}$ with  coupling strength $g$. The rotating-wave approximation is assumed for $H_{\text{int}}$  to conserve the total energy. 

The emitter-cavity array can exhibit single excitation bound state, for which one excitation is confined around the emitters, resulting in radiation  suppression \cite{john91,kofman94,calojo,marinica,bulgakov,plotnik}.   In this model, two types of bound states can be identified: discrete bound states (DBS) with an energy gap from the continuum enabling photon preservation against spontaneous emission \cite{john91,kofman94,calojo},   and  bound states in continuum (BIC) existing within the continuum due to destructive interference between different channels, leading to complete radiation elimination  \cite{marinica,bulgakov,plotnik}.  These bound states can trigger an effective interaction among emitters \cite{calojo}.  The nature of the effective interaction, whether exponentially decaying or sinusoidally oscillating, is determined by the presence of DBS or BIC, ultimately influencing the localization characteristics of the bound states.  As a consequence, the dynamics of single excitation in emitters is solely governed by these bound states, serving as the focal point of the ensuing discussion.

\emph{Determination of bound states} Due to the absence of particle interaction, the study is limited to the single excitation subspace. Consequently, the eigenstate can be written formally as  
$\ket{\psi}=\left( \sum_{n=1}^{N}\alpha_n \sigma^{+}_n + \sum_{j=1}^{N} \beta_j b^{\dagger}_j \right)\ket{0}$, 
where $\ket{0}$ denotes the vacuum. Substituting the relation above into Sch\"{o}dinger equation, one obtains a linear system of equations for $\alpha_n$ or $\beta_n$,
\be 
\left(H_s - E \right)\overline{\alpha}+H_{\text{int}}\overline{\beta}&=& 0, \nonumber \\
H^{\dagger}_{\text{int}}\overline{\alpha}+ \left(H_b - E \right)\overline{\beta}&=& 0
\ee
where $\overline{\alpha}=\left(\alpha_1, \alpha_2, \cdots, \alpha_N \right)^{T}, \overline{\beta}= \left(\beta_1, \beta_2, \cdots, \beta_N \right)^{T}$. Through eliminating $\overline{\beta}$ \cite{hu25a,hu25b}, one gets 
\be \label{Hbareqn}
\overline{H}_s\overline{\alpha}= E \overline{\alpha},
\ee
where  $\overline{H}_s$ denotes the effective Hamiltonian  for emitters, defined as  
\be \label{Hbar}
\overline{H}_s=H_s  + H_{\text{int}} \frac{1}{E - H_b} H^{\dagger}_{\text{int}}.
\ee
Notably, $\overline{H}_s$ is involved with $E$, which makes the determination of eigenvalue nontrivial. To determine $E$ formally, one must solve the equation $\det \left(\overline{H}_s -E \right)=0$. Subsequently, $\overline{\alpha}$ can be determined  by substituting $E$ back into Eq. \eqref{Hbareqn} and  identifying the eigenstate of $\overline{H}_s$ with the corresponding eigenvalue. It should be noted that  the Hermiticity of $\overline{H}_s$ can be maintained  only for real values of $E$, highlighting the stability of excitations in emitters.  Moreover,  the quantity of real $E$s typically exceeds $N$, thereby introducing a higher level of complexity in the dynamics of excitations.  

For a real value of $E$, the corresponding $\overline{\alpha}$ represents the photon-emitter bound state, facilitating the evaluation of localization of single excitation within the emitter system. It is essential to acknowledge that these bound states are typically non-orthogonal, precluding them from constituting a complete set of eigenstates that characterize excitation dynamics in emitters. This characteristic arises from the elimination of cavity degrees of freedom, diminishing the impact of cavities as the effective interaction, as shown by the second term in Eq. \eqref{Hbar}. Nevertheless, the bound states depicted by $\overline{\alpha}$s,  govern the steady-state characteristics of excitation in emitters.  Consequently, the subsequent discourse focuses on identifying  real values of $E$ and assessing localization within emitters by analyzing $\overline{\alpha}$.

To determine the real value of $E$ and its corresponding bound state $\overline{\alpha}$, it is essential to find the specific form for the second  term in Eq. \eqref{Hbar}. This term exhibits  divergence at $E=\omega_k$, necessitating evaluation in two distinct regions: $\left|E-\omega_c \right|>2J $ or $\left|E-\omega_c \right|<2J $. As shown in supplement material (SM) \ref{SM:Hs}, DBS emerges for the former, while BIC arises for the latter. Consequently, the interaction term can manifest in three distinct configurations,
\begin{widetext}
\be\label{Hint}
\overline{H}_{\text{int}}= H_{\text{int}} \frac{1}{E - H_b} H^{\dagger}_{\text{int}}=\frac{g^2}{2J}\sum_{n, n'} \sigma^{+}_n \sigma^{-}_{n'}\times \begin{cases}
\displaystyle
- \frac{\left(-1\right)^{\left|n-n'\right|}}{\sinh x} e^{-\left|n-n'\right| x }, & \mbox{if } \displaystyle \left|\frac{\omega_c -E }{2J}\right|>1 \mbox{ and }   E< \omega_c \\
\displaystyle\frac{1}{\sinh x} e^{-\left|n-n'\right| x }, & \mbox{if } \displaystyle \left|\frac{\omega_c -E }{2J}\right|>  1 \mbox{ and }  E> \omega_c \\
 \displaystyle-\frac{1}{\sin \theta} \sin \left(\left|n-n'\right| \theta \right), & \mbox{if }\displaystyle \left|\frac{\omega_c -E }{2J}\right|< 1, 
  \end{cases}
\ee
\end{widetext}
where $\cosh x = \left|\frac{E- \omega_c}{2J}\right|, \cos\theta=  \frac{E- \omega_c}{2J}$,
and the symbol $\overline{H}_{\text{int}}$ is used for brevity.  Evidently, an energy-dependent long-range correlation can be triggered in  emitters through their coupling to cavities. Especially, the correlation can exhibit  exponentially decaying or sinusoidally oscillating with distance, dependent on the occurrence of DBS or BIC. This difference can lead to  the distinct localization characters of excitation within emitters. 

\begin{figure*}
\center
\includegraphics[width=16cm]{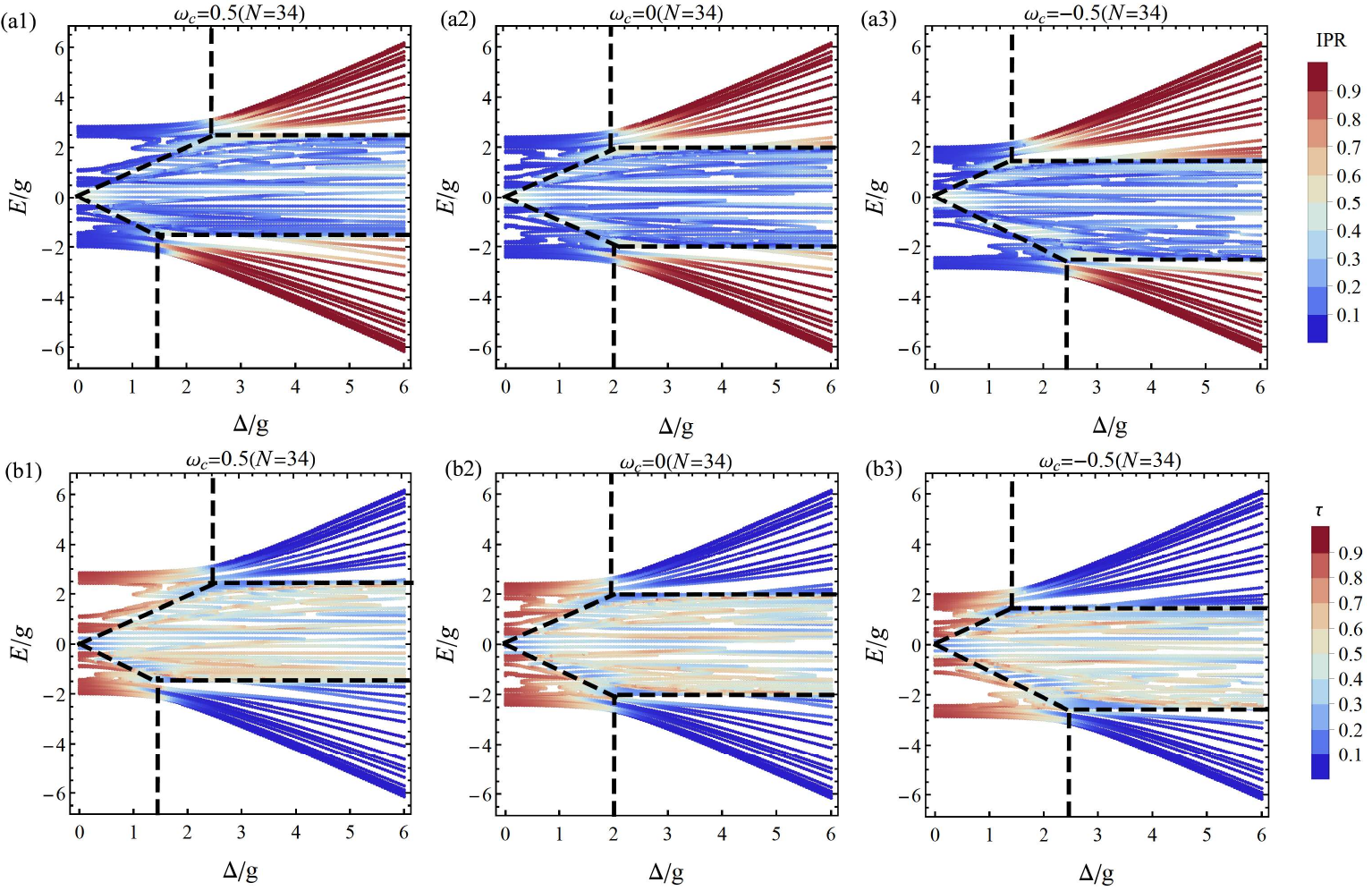}
\caption{(Color online) Comparative diagrams of the  inverse participation ratio (a1)-(a3) and the fractal dimension (b1)-(b3) are presented for the bound state function $\overline{\alpha}$s under $\omega_c=0.5, 0, -0.5$ respectively. The dashed-black lines represent the mobility edges $E_c$ or the critical values  $\Delta_c$, defined by Eq. \eqref{MEs}. In all figures, $E$ and $\Delta$ are scaled in units of $g$.  All numerical calculations adopt fixed parameters $N=34$ and $g=1$.  }
\label{fig:IPRandFD}
\end{figure*}

\emph{Localization character of bound states} 
To quantify the localization of bound states, we adopt the inverse participation ratio (IPR) and the fractal dimension (FD) $\tau$, which are defined respectively as 
\be
\text{IPR}&=&\sum_{n=1}^{N} \left|\alpha_n\right|^4, \\
\tau&=& -\frac{\ln \text{IPR}}{\ln N}.
\ee
For localized states, $\text{IPR}$ is unity, whereas  it approaches zero for extended state. Conversely, $\tau$ is unity for extended state, but is zero for localized state.  The intermediate value of either IPR or $\tau$ typically defines the critical state \cite{zhou23}. The determined  bound-state levels are  depicted as functions of  $\Delta$ in Fig. \ref{fig:IPRandFD}, accompanied by corresponding IPR (a1-a3) or $\tau$ (b1-b3) values.  It is manifest the $E-\Delta$ plane reveals distinct regions based on the IPR or $\tau$ values, indicating extended, localized, or critical characteristics of the bound states. In SM \ref{SM:ME}, four MEs and a critical $\Delta$  can be identified analytical through a generalized duality transformation, which are written as 
\be\label{MEs} 
E_c &=& \omega_c \pm 2J, \nonumber \\
E_c&=& \pm \Delta , \\
\Delta_c &=& \pm \omega_c + 2J.\nonumber
\ee
As illustrated in Fig. \ref{fig:IPRandFD}, $E_c$s and $\Delta_c$ notably characterize the boundaries of these regions, as denoted by the dashed-black lines.  

Some comments are in place for the observations shown in Fig. \ref{fig:IPRandFD}. It is noted first for $\Delta > \pm\omega_c +2J$ that the localized states can be found in region $\left|\omega_c -E \right|>  2J$, where only DBS occurs. In contrast, critical states can be found in region $\left|\omega_c -E \right|< 2J$, where BIC occurs solely. This phenomenon can be attributed to the unique long-range correlations displayed by Eq. \eqref{Hint}. Notably, an exponentially decaying hopping of excitation among emitters is induced when $\left|\omega_c -E \right|>  2J$ is satisfied. In such case, as increment of $\left| E \right|$, not only  is the hopping strength $\frac{1}{\sinh x}$ reduced, but also dose the correlation length depicted by $1/x$ decreases rapidly. Together with the strong  quasiperiod depicted by a large $\Delta$, the hopping of excitation among emitter is compressed significantly. On the contrary, a sinusoidally oscillating hopping can emerge if $\left|\omega_c -E \right|< 2J$ is satisfied, as shown in Eq. \eqref{Hint}. Consequently, the excitation hopping among emitters can persists regardless of the quasiperiodicity.  Therefore, the critical state arises from the  competition between the  quasiperiodic potential and the  sinusoidally oscillating hopping. 

The situation changes when $\Delta < \pm\omega_c +2J$. In such case,  the mobility edge $E_c=\pm \Delta$ divide the  $E-\Delta$ plane into two distinct regions.  For $\left| E \right| <\Delta$, where only BIC occurs, the bound states display critical behavior. In contrast, for $\left| E \right| >\Delta$, where both DBS and BIC coexist, the bound states becomes extended.  This picture can be explained by the competition between  the onsite potential $\Delta_n$ and the $E$-related hopping. It is worth noting that  $\Delta < \pm\omega_c +2J$ does not exceed the bandwidth of the cavity continuum; therefore,  single excitation may acquire energy from to overcome the trapping of quasiperiodic potential $\Delta_n$ when $\left| E \right| > \Delta$. On the other hand, when $\left| E \right| < \Delta$, the interplay between the onsite potential $\Delta_n$ and $E$-related hopping dominates the dynamics of single excitation, giving rise to the observed critical behavior.

\emph{Influence of the edge state} However, it is noteworthy that the localization of bound states near $E=0$ deviates markedly from the predication of Eq. \eqref{MEs} at small values of $\Delta$. As illustrated in SM \ref{SM:edgestate}, this discrepancy stems from the emergence of an edge state under the open boundary condition for $\Delta=\omega_c=0$ and $N=2\overline{N}+1$ ($\overline{N}$ being integer number), i.e. 
$\ket{E=0}= \frac{1}{\sqrt{2}} \left[ \sigma^{\dagger}_1 + \left( -1 \right)^{\overline{N}} \sigma^{\dagger}_N \right] \ket{0}$. It is clear that the excitation of $\ket{E=0}$ can be located uniquely at the end points with identical probability. As the energy $E$ deviates from zero,  the excitation can be populated near the end points with exponentially decaying amplitude (see Fig. \ref{figS:edgestate} in SM. \ref{SM:edgestate}). Nonzero $\Delta$ or $\omega_c$ may destroy the edge state since the zero-energy mode no longer constitutes a steady state of the system. Accordingly,  Fig. \ref{fig:IPRandFD} distinctly exhibits  that the deviation vanishes  with the increase of $\Delta$.   

\begin{figure*}
\center
\includegraphics[width=\textwidth]{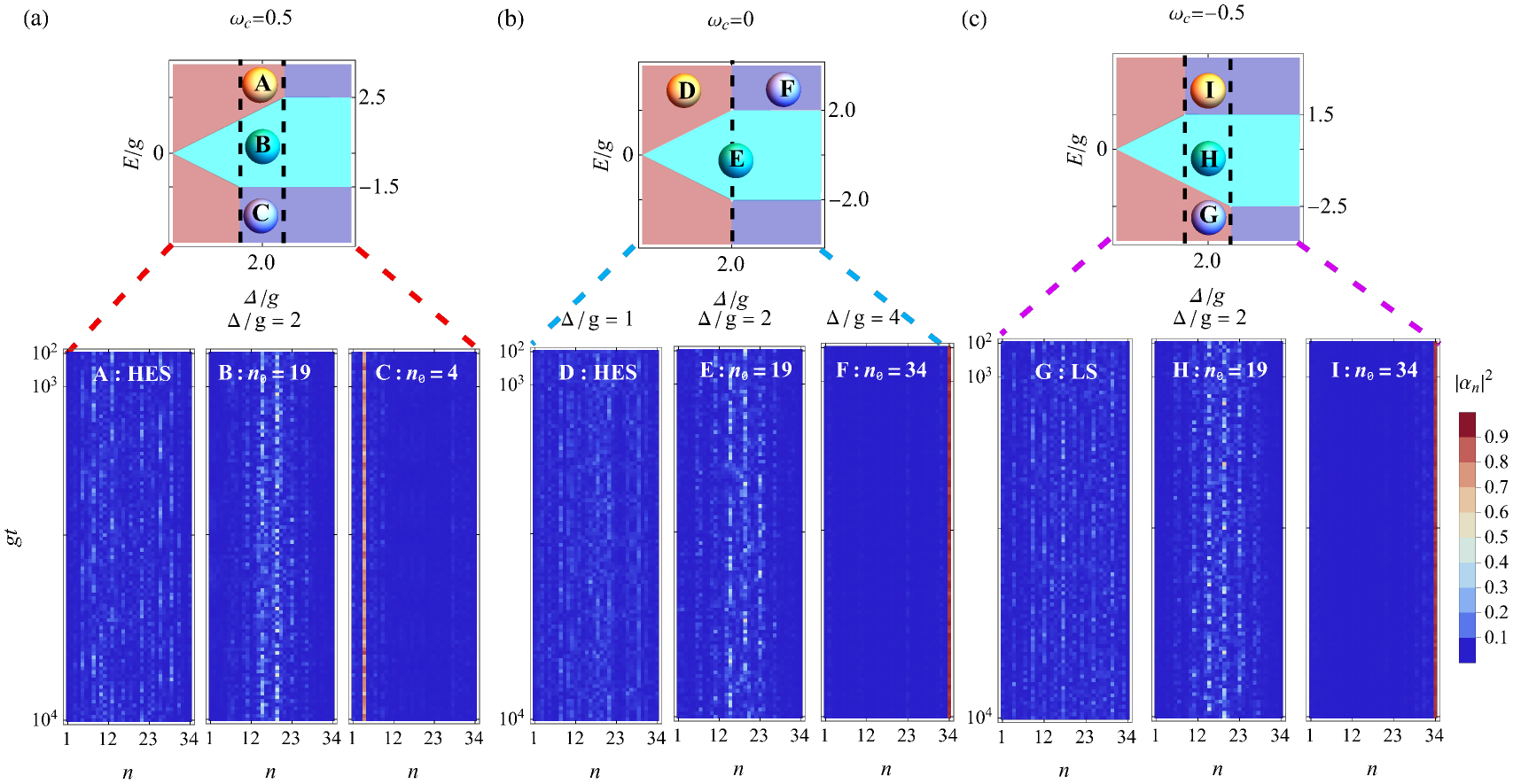}
\caption{(Color online) The long-term temporal evolution of $\alpha_n(t)$ for selected initial states, located initially at the extended, localized and critical regions in the phase diagram.  The value of $n_0=4, 19, 34$ denotes the initial site of excitation in emitters, where $\Delta_4/\Delta=-0.985,\Delta_{19}/\Delta=-0.046$ and $\Delta_{34}/\Delta=0.997 $.  Thus, the value of $\Delta$ decides the position of initial state in the phase diagram. The "HES" denotes the highest excited bound state, while "LS" denotes the lowest bound state.  $N=34$ and $g=1$ are chosen for all plots.    }
\label{fig:evolution}
\end{figure*}

\emph{Long-time dynamics of single excitation} To verify the phase diagram shown in Fig. \ref{fig:phase}(b), it is necessary to find the temporal evolution for single excitation in emitters. By Schr\"{o}dinger equation, one obtains
\be 
\text{i}\frac{\partial \alpha_n(t) }{\partial t} &=& \Delta_n  \alpha_n(t)- \text{i}\frac{g^2}{2\pi}\sum_m \int_{-\pi}^{\pi} \text{d}k e^{-\text{i}\left(n-m\right)k} \times \nonumber \\
&& \int_{0}^{t} \text{d}s \alpha_m(s) e^{- \text{i} \left( \omega_c + 2 J \cos k \right)\left(t-s\right) }
\ee   
To disclose the long-time dynamics of $\alpha_n(t)$,  Laplace transformation $A_n(E)= \int_{0}^{\infty} \text{d} t e^{- \text{i} E t} \alpha_n(t)$ is imposed. Resultantly, we have  
\be 
\left( \Delta_n - E\right) A_n(E) +\frac{g^2}{2\pi} \sum_m A_m(E) \times \nonumber \\
\int_{-\pi}^{\pi} \text{d}k \frac{ e^{-\text{i}\left(n-m\right)k}  }{E-\left( \omega_c + 2 J \cos k \right)}=- \text{i} \alpha_n(0).
\ee
Thus, $\alpha_n(t)$ can be decided  through inverse Laplace transformation, 
\be 
\alpha_n(t)=\frac{1}{2\pi } \int_{\text{i}s-\infty}^{\text{i}s+\infty} \text{d}E A_n(E) e^{- \text{i} E t}
\ee
By residue theorem, one obtains 
\be 
\alpha_n(t)= \sum_j c_{n,j}e^{- \text{i}E_j t}, 
\ee
where $c_{n,j}$ denotes the residue of $A_n(E)$ at the bound-state level $E_j$.

Figure \ref{fig:evolution} shows the long-time evolution of $\left|\alpha_n(t)\right|^2$ for initial states with different energies Evidently,  $\left|\alpha_n(t)\right|^2$ can display extended, localized or critical behavior,  decided fully  by  the initial state’s position in the phase diagram shown in  Fig. \ref{fig:phase}(b). For the cases illustrated in Fig. \ref{fig:evolution} A, D and G, $\left|\alpha_n(t)\right|^2$ tends to be uniform across all sites, reflecting that the initial states lie within the extended phase. In contrast,  when the excitation is initially located at site $n=19$, $\left|\alpha_n(t)\right|^2$ remains predominantly confined to a few sites around $n=19$, as seen in  Fig. \ref{fig:evolution} B, E and H. This behavior arises because the corresponding initial states have energies close to zero and therefore reside in the critical phase.

Interestingly, the central frequency $\omega_c$ can strongly influence the localization properties of  $\left|\alpha_n(t)\right|^2$. As For $\omega_c=0.5$, as shown in Fig. \ref{fig:evolution} C,  the excitation remains predominantly localized at the initial site  $n=4$ since the corresponding initial state has low energy of  $\sim -2$ when $\Delta /g=2$, placing it in the localized phase. On the contrary, the situation is reversed  for the same  $\Delta /g=2$ when $\omega_c=-0.5$. As shown in Fig. \ref{fig:evolution} I, the excitation initially located at site $n=34$ exhibits pronounced localization. In such case, the corresponding initial state possesses a  high energy of  $\sim +2$, again lying within the localized phase. These results demonstrate that the spatial localization of excitations in emitters   can be effectively controlled by tuning the cavity field via $\omega_c$.  


Finally, we emphasize   that the time evolution of  $\left|\alpha_n(t)\right|^2$ cannot be expressed as a coherent superposition of bound states with  dynamical phase factor $e^{-\text{i}E_i t}$.  The reason is because the energy  $E$ enters the definition of $\overline{H}_s$ as shown  by  Eq. \eqref{Hbar},  rendering the resulting eigenfunctions nonorthogonal. Thus, bound states alone do not provide a complete description of the  excitation dynamics.  Actually, the definition of $\overline{H}_s$ permits complex $E$ with negative imaginary part, which describes  the dissipation of excitation \cite{cui25}. These complex energy levels are omitted deliberately here, as they do not contribute to the long-time stable dynamics of excitation in emitters. 

\emph{Experimental proposal} The physical realization of the configuration illustrated in Fig. \ref{fig:phase}(a) remains experimentally challenging. A key difficulty lies in the precise positioning of each emitter within its respective cavity. Furthermore, measuring the excitation population $\left|\alpha_n(t)\right|^2$ requires individual readout of individual emitter, which is time-consuming and becomes increasingly impractical for large system sizes $N$ \cite{wang25}. Recently, a cavity-array microscope has been proposed as an experimental platform that integrates emitter arrays with optical cavities in a parallelized and scalable manner \cite{shaw26}.This architecture is naturally compatible with existing emitter-array technologies and can be readily integrated into such systems. Owing to these advantages, the scheme discussed here becomes experimentally feasible.

The long-time dynamics of single excitation can be accessed under the photon-blockade regime,  where only a single excitation is allowed in the emitter–cavity system \cite{imamoglu}. In this regime, the population $\left|\alpha_n(t)\right|^2$  can be measured by imaging the  emitters with single-site resolution. The imaging signal shows distinct loss feature decided by the differential light shift of the ground and excited  states of the emitters \cite{shaw26}.     

\emph{Conclusion and outlook} In conclusion, we have presented a comprehensive study of  single excitation localization in a emitter-cavity arrays subjected to a quasiperiodic potential. By identifying  emitter-photon bound states,   we construct an effective single-excitation Hamiltonian $\overline{H}_s$ for different energy regimes. The Hamiltonians characterize exponentially decayed or sinusoidally oscillated hopping of excitation among emitters, giving rise to the  extended, localized and critical bound states. The resulting localization behavior is fully determined by the interplay between the quasiperiodic potential strength $\Delta$,  the cavity central frequency $\omega_c$ and the bound state energy $E$. These relationships are established analytically via a duality transformation.We identify four mobility edges and a critical quasiperiodic strength $\Delta_c$, which together yield the phase diagram  shown in Fig. \ref{fig:phase}(b). To  verify the phase diagram,  we examine the long-time dynamics of  for the initial states located at the different regimes of phase diagram. As demonstrated in Fig. \ref{fig:evolution}, the dynamical evolution is in full agreement with the predicted localization phases.

An important feature of the phase diagram  in Fig. \ref{fig:phase} (b) is that the cavity central frequency $\omega_c$ strongly influences  localization of single excitation. This clearly demonstrates that the excitation dynamics can be controlled by tuning the cavity field. Accordingly, emitter–cavity arrays offer a versatile platform for manipulating excitation transport.    

An open question is how the Kerr nonlinearity, i.e. physics beyond the single excitation subspace, affect single-excitation localization.  Recent studies indicate that particle interaction can significantly modify the single-excitation MEs,with the precise effect being highly model dependent \cite{modak15,nag17,alexan18,kohlert,wei19,alexan21,wangyunfei21,huang23,huang24,huang24}. This suggests that  the influence of interaction is sensitive to the specific  physical properties of the system. In the present  model, the photon interaction depicted by Kerr nonlinearity enhances photon hopping between cavities,leading us to conjecture that the phase diagram will be modified. Moreover, beyond single-excitation physics, the emergence of multi-excitation bound states \cite{winkler06} raises the intriguing possibility of the many-body mobility edges. These questions lie beyond the scope of this work and will be addressed in future studies. 

\emph{Acknowledgement} The author (H. T. Cui) acknowledges the support of Provincial Natural Science Foundation (ZR2021MA036).  H. Z. Shen acknowledges the supports of Science and Technology Development Plan Project of Jilin Province (20250102007JC) and National Natural Science Foundation of China (12274064). X. X. Yi acknowledges the support of National Natural Science Foundation of China (12575010).

\clearpage

\newpage



\onecolumngrid

\setcounter{page}{1}

\setcounter{figure}{0}                     
\renewcommand{\thefigure}{S\arabic{figure}}

\renewcommand{\theequation}{S\arabic{equation}}
\setcounter{equation}{0}

\section*{Supplemental Material for ``Dissipation induced anomalous mobility edges in a quasidisordered atom-cavity  chain"}
\begin{minipage}[c]{\textwidth}
\center{H. T. Cui$^{1}$,  H. Z. Shen $^{2}$,   M. Qin$^{1}$,  and X. X. Yi $^{2}$ \\
\it{ \small $^1 $ School of Physics and Optoelectronic Engineering \and Institute of Theoretical Physics, Ludong University, Yantai 264025, China and \\
 $^2$ Center for Quantum Sciences, Northeast Normal University, Changchun 130024, China}} \\
(\today)
\end{minipage}

\subsection{Evaluation of the item $H_{\text{int}} \frac{1}{E - H_b} H^{\dagger}_{\text{int}}$}
\label{SM:Hs}

To evaluate $H_{\text{int}} \frac{1}{E - H_b} H^{\dagger}_{\text{int}}$ in Eq. \eqref{Hbar}, it is convenient to introduce the Fourier transformation in the momentum space,
\be 
b_j= \frac{1}{\sqrt{N}} \sum_k e^{- \text{i} \frac{2\pi k}{N} j }b_k, b^{\dagger}_j= \frac{1}{\sqrt{N}} \sum_k e^{\text{i} \frac{2\pi k}{N} j }b^{\dagger}_k.
\ee
Then,  $H_b$ can be diagonalized as 
\be 
H_b=\sum_{k=-\left[N/2\right]}^{\left[N/2\right]} \left(\omega_c + 2J \cos\frac{2\pi k}{N} \right)b^{\dagger}_k b_k,
\ee
with eigenvector $\ket{k}$ for a definite $k$ written by
\be 
\ket{k}=\frac{1}{\sqrt{N}} \left( e^{ \text{i} \frac{2\pi k}{N}},  e^{ \text{i} \frac{4\pi k}{N}}, \cdots, e^{ \text{i} 2\pi k} \right)^T.
\ee
The transformation matrix between the space and momentum representations can be constructed as
\be 
U=\left(\ket{k=-\left[N/2\right]}, \ket{k=-\left[N/2\right]+1}, \cdots, \ket{k=\left[N/2\right]} \right)\Rightarrow U_{j, k}=  e^{ \text{i} \frac{2\pi k}{N} j}.
\ee
We thus obtain 
\be 
&& \frac{1}{E - H_b} = U U^{\dagger } \frac{1}{E - H_b}U U^{\dagger }=U \text{Diagonal}\left(\cdots, \frac{1}{E- \left(\omega_c + 2J \cos\frac{2\pi}{N} k \right)}, \cdots \right) U^{\dagger }\nonumber \\
 &&\Rightarrow \left( \frac{1}{E - H_b} \right)_{j, j'}= \frac{1}{N} \sum_k  \frac{e^{\text{i}\frac{2\pi k}{N} \left(j-j' \right) }}{E- \left(\omega_c + 2J \cos\frac{2\pi k}{N} \right)}.
\ee
Noted that  the matrix element for $H_{\text{int}}$ can be written in the space representation with single excitation as 
\be 
\left(H_{\text{int}} \right)_{n,j}=g \delta_{n,j} \rightarrow \left(H^{\dagger}_{\text{int}} \right)_{j, n}= g \delta_{j,n}, 
\ee
we find 
\be 
\left(H_{\text{int}} \frac{1}{E - H_b} H^{\dagger}_{\text{int}} \right)_{n,n'}&=& \sum_{j, j'} \left(H_{\text{int}} \right)_{n,j} \left( \frac{1}{E - H_b} \right)_{j, j'} \left(H^{\dagger}_{\text{int}} \right)_{j', n}= g^2  \sum_{j, j'} \delta_{n,j}  \left( \frac{1}{E - H_b} \right)_{j, j'} \delta_{j',n}=g^2 \left( \frac{1}{E - H_b} \right)_{n, n'} \nonumber \\ 
&=&\frac{g^2}{N} \sum_k \frac{e^{\text{i}\frac{2\pi k}{N} \left(n-n' \right) }}{E- \left(\omega_c + 2J \cos\frac{2\pi k}{N} \right)}.
\ee

For  $N\rightarrow \infty$, $\frac{1}{N} \sum_k$ is replaced by the integral $\frac{1}{2\pi} \int_{-\pi}^{\pi} \text{d}k$. Resultantly, one finds 
\be 
\left(H_{\text{int}} \frac{1}{E - H_b} H^{\dagger}_{\text{int}} \right)_{n,n'}= \frac{g^2}{2\pi} \int_{-\pi}^{\pi} \text{d}k  \frac{e^{\text{i}k \left(n-n' \right) }}{E- \left(\omega_c + 2J \cos k \right)},
\ee
which is invariant  under exchanging $n\leftrightarrow n'$. The integral above can be evaluated exactly  when it is transformed into a contour integral along a unity circle in the complex $k$-plane. Defining $z=e^{\text{i}k}$, we find 
\be 
\frac{g^2}{2\pi} \int_{-\pi}^{\pi} \text{d}k  \frac{e^{\text{i}k \left(n-n' \right) }}{E- \left(\omega_c + 2J \cos k \right)} \Rightarrow \frac{g^2}{2\pi \text{i}} \oint_{\left|z\right|=1} \frac{z^{\left|n-n'\right|}}{-Jz^2 + \left( E-\omega_c \right)z-J  },
\ee
which can be evaluated exactly  by residual theorem. The singular points satisfy equation
\be\label{zeqn} 
Jz^2 - \left( E-\omega_c \right)z+J=0,
\ee
which obviously can be solved for two distinct cases. 
\begin{enumerate}
  \item  Case A: $\displaystyle \left|\frac{E- \omega_c}{2J}\right|> 1$
 
In this situation, there are two roots for Eq. \eqref{zeqn}, i.e.
\be 
z_{\pm}=\frac{E- \omega_c}{2J} \pm \sqrt{\left( \frac{E- \omega_c}{2J}\right)^2-1}, 
\ee  
which satisfies relation $z_+ z_-=1$. For convenience, we introduce the following definition 
\be 
\cosh x=\left|\frac{E- \omega_c}{2J}\right|, \sinh x=\sqrt{\left( \frac{E- \omega_c}{2J}\right)^2-1}. 
\ee
It is obvious for $E<\omega_c$ that $\left|z_- \right|>1$ and the singular point is just $z_+$. By residual theorem, one has
\be 
 \frac{g^2}{2\pi \text{i}} \oint_{\left|z\right|=1} \frac{ z^{\left|n-n'\right|}}{-Jz^2 + \left( E-\omega_c \right)z-J  }= - \frac{g^2z_+^{\left|n-n'\right|}}{J\left(z_+-z_- \right) }=- \frac{g^2\left(-1\right)^{\left|n-n'\right|}}{2J\sinh x} e^{-\left|n-n'\right| x }.
\ee
While for  $E>\omega_c$, $\left|z_+ \right|>1$,  $z_-$ is singular point. We has
\be 
 \frac{g^2}{2\pi \text{i}} \oint_{\left|z\right|=1} \frac{z^{\left|n-n'\right|}}{-Jz^2 + \left( E-\omega_c \right)z-J  }= - \frac{g^2z_-^{\left|n-n'\right|}}{J\left(z_--z_+ \right) } = \frac{g^2}{2J\sinh x} e^{-\left|n-n'\right| x }.
\ee

\begin{figure}
\center
\includegraphics[width=4cm]{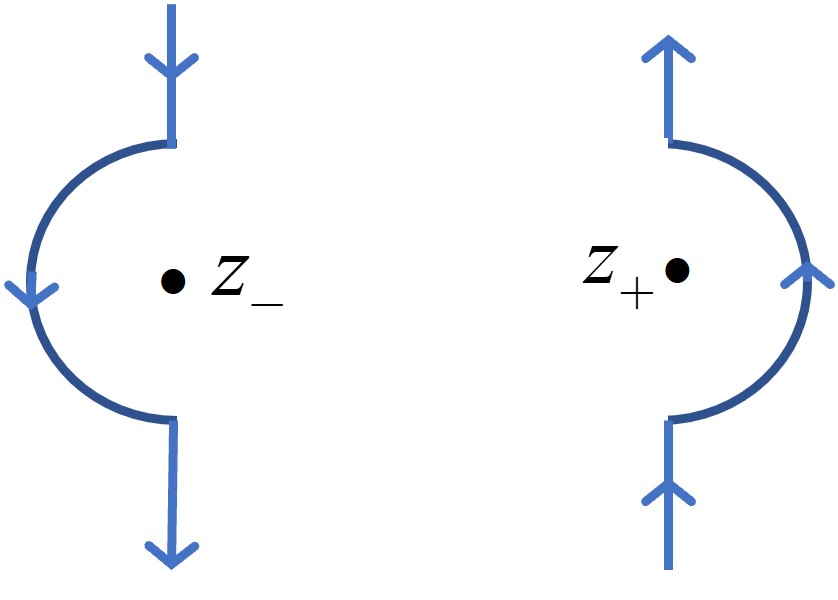}
\caption{ The deformed contour around $z_{\pm}=e^{\pm\text{i}\theta}$ in order to evaluate  Eq. \eqref{zeqn}.  }
\label{figS:contour}
\end{figure}

\item Case B: $\displaystyle \left|\frac{\omega_c -E }{2J}\right| \leq 1$

The two roots can be written as
\be 
z_{\pm}=\frac{E- \omega_c}{2J} \pm \text{i}\sqrt{1- \left( \frac{E- \omega_c}{2J}\right)^2}.
\ee
Defining 
\be 
\cos \theta= \frac{E- \omega_c}{2J}, \sin\theta=\sqrt{1- \left( \frac{E- \omega_c}{2J}\right)^2},
\ee
one can find $z_{\pm}=e^{\pm\text{i}\theta}$. Since the two roots are located on the contour, one can deform the contour around $z_{\pm}$, as shown in Fig. \ref{figS:contour}. Consequently, one has 
\be 
\frac{g^2}{2\pi \text{i}} \oint_{\left|z\right|=1} \frac{z^{\left|n-n'\right|}}{-Jz^2 + \left( E-\omega_c \right)z-J  }= -\frac{g^2}{2J}\left(\frac{z_+^{\left|n-n'\right|}}{z_+ - z_-} +\frac{z_-^{\left|n-n'\right|}}{z_- - z_+}  \right)= -\frac{g^2}{2J \sin \theta} \sin \left(\left|n-n'\right| \theta \right) 
\ee
\end{enumerate}

Finally, the effective hamiltonian can be written explicitly as 
\be\label{effectHs} 
\overline{H}_s= \sum_{n=1}^{N} \Delta_n \sigma^{+}_n \sigma^{-}_{n} + \frac{g^2}{2J}\sum_{n, n'} \sigma^{+}_n \sigma^{-}_{n'}\times \begin{cases}
\displaystyle
- \frac{\left(-1\right)^{\left|n-n'\right|}}{\sinh x} e^{-\left|n-n'\right| x }, & \mbox{if } \displaystyle \left|\frac{\omega_c -E }{2J}\right|>1 \mbox{ and }   E< \omega_c \\
\displaystyle\frac{1}{\sinh x} e^{-\left|n-n'\right| x }, & \mbox{if } \displaystyle \left|\frac{\omega_c -E }{2J}\right|>  1 \mbox{ and }  E> \omega_c \\
 \displaystyle-\frac{1}{\sin \theta} \sin \left(\left|n-n'\right| \theta \right), & \mbox{if }\displaystyle \left|\frac{\omega_c -E }{2J}\right|< 1.
 \end{cases}
\ee

\subsection{Determination of mobility edges}  \label{SM:ME}
As for the long-range hopping among emitters shown in Eq. \eqref{effectHs}, we adopt the same approach as done in Ref.\cite{biddle} to determine  mobility edges. However, noting that the forms of the long-range hopping is strongly correlated to the ratio between $\omega_c - E$ and $2J$, as well as $E>\omega_c$ or not. The discussion in this subsection is also divided into three parts.

\subsubsection{Case A: $ \displaystyle \left|\frac{\omega_c -E }{2J}\right|>  1$ and  $E> \omega_c$}
By Eq. \eqref{Hbareqn}, one can obtain
\be\label{alpha1} 
\Delta \left(\cos 2\pi \beta n - \frac{E}{\Delta} \right)\alpha_n + \frac{g^2}{2J \sinh x} \sum_{n'}e^{-\left|n-n'\right| x } \alpha_{n'} =0.
\ee
In this case, we introduce the generalized duality transformation defined as 
\be\label{T1}
\alpha_k&=&\sum_n e^{\text{i}2\pi \beta n k} T_n(x_0) \alpha_n,  T_n(x_0) \alpha_n =\sum_k e^{- \text{i}2\pi \beta n k} \alpha_k. \\
T^{-1}_n(x_0)&=& \sum_{m= -\infty}^{+\infty} e^{-x_0 \left| m \right|}  e^{\text{i}2\pi \beta n m} =\frac{\sinh x_0}{\cos 2\pi \beta n -\cosh x_0},
\ee   
where $\cosh x_0 =\frac{E}{\Delta}$. Notably, 
\be
T_n(x_0) = T_{-n}(x_0), \lim_{n\rightarrow \infty}\sum_n  e^{\text{i}2\pi \beta n \left( m -m' \right)}=\delta\left( m -m'  \right). 
\ee

Noting that 
\be 
\Delta \left(\cos 2\pi \beta n - \frac{E}{\Delta} \right)\alpha_n= \Delta \sinh x_0  T_n(x_0) \alpha_n,
\ee 
and utilizing Eq. \eqref{T1}, we have 
\be
\Delta \sum_n e^{\text{i}2\pi \beta n k}  \left(\cos 2\pi \beta n - \frac{E}{\Delta} \right)\alpha_n = \Delta \sinh x_0 \sum_n e^{\text{i}2\pi \beta n k} T_n(x_0) \alpha_n= \Delta \sinh x_0  \alpha_k. 
\ee 
Similarly, one can find
\be
\sum_{n'}e^{-\left|n-n'\right| x } \alpha_{n'}&=& \sum_{n'}e^{-\left|n-n'\right| x } T^{-1}_{n'}(x_0) \sum_{k'} e^{- \text{i}2\pi \beta n' k'} \alpha_{k'}   \nonumber \\
&=&  \sum_{k'} \alpha_{k'} \left(\sum_m e^{-x_0 \left| m \right|}  e^{\text{i}2\pi \beta n' m} \right) \sum_{n'} e^{-\left|n-n'\right| x } e^{- \text{i}2\pi \beta n' k'}\nonumber \\
&=&  \sum_{k', m }\alpha_{k'}  e^{-x_0 \left| m \right|} e^{\text{i}2\pi \beta n\left( m- k'\right)} \left[\sum_{n'}e^{-\left|n-n'\right| x }   e^{\text{i}2\pi \beta \left( m- k'\right)\left(n'-n \right)} \right] \nonumber \\
&=&   \sum_{k', m }\alpha_{k'}  e^{-x_0 \left| m \right|} e^{\text{i}2\pi \beta n\left( m- k'\right)}  T^{-1}_{m-k'}\left(x\right).
\ee
Multiplying the above equation by $\sum_n e^{\text{i}2\pi \beta n k} $, one gets
\be 
&\Rightarrow & \sum_{k', m }\alpha_{k'}  e^{-x_0 \left| m \right|}   T^{-1}_{m-k'}\left(x\right) \sum_n e^{\text{i}2\pi \beta n\left( m- k'+k\right)} \nonumber \\
&=&   T^{-1}_{k}\left(x\right)  \sum_{k' }\alpha_{k'}  e^{-x_0 \left| k-k' \right|},  
\ee
for which $\sum_n e^{\text{i}2\pi \beta n\left( m- k'+k\right)} = \delta\left(m- k'+k \right) $ is used. 

Finally, one transform Eq. \eqref{alpha1} into the duality momentum space
\be
 \Delta \sinh x_0  \alpha_k + \frac{g^2}{2J \sinh x} T^{-1}_{k}\left(x\right)  \sum_{k' }\alpha_{k'}  e^{-x_0 \left| k-k' \right|} =0 \nonumber \\
 \label{alpha1k}
\Rightarrow \Delta \sinh x_0 T_{k}\left(x\right) \alpha_k+ \frac{g^2}{2J \sinh x} \sum_{k' }\alpha_{k'}  e^{-x_0 \left| k-k' \right|}=0. 
\ee
Evidently when $\cosh x =\cosh x_0 $ and thus $T_{k}\left(x\right)=T_{k}\left(x_0\right)$,  Eq. \eqref{alpha1k} and Eq. \eqref{alpha1} can be transformed into each other by exchanging subscripts $n$ and $k$, which means that the system of emitters becomes dual invariant. Noted that both  $x_0$ and $x$ are related to $E$, the mobility edge termed as $E_c$ thus satisfies 
\be 
\frac{E_c- \omega_c}{2J} =\frac{E_c}{\Delta}.
\ee  

\subsubsection{Case B: $ \displaystyle \left|\frac{\omega_c -E }{2J}\right|>  1$ and  $E< \omega_c$}
In this situation, one can find by Eq. \eqref{Hbareqn}
\be\label{alpha2} 
\Delta \left(\cos 2\pi \beta n - \frac{E}{\Delta} \right)\alpha_n - \frac{g^2}{2J \sinh x} \sum_{n'} \left(-1\right)^{\left|n-n'\right|} e^{-\left|n-n'\right| x } \alpha_{n'} =0.
\ee
Adopting the similar approach above, we define $T_n\left( x_0 \right)$ as
\be
T^{-1}_n\left( x_0 \right)= \sum_{m= -\infty}^{+\infty} e^{-x_0 \left| m \right|} \left(-1\right)^{\left|m\right|} e^{\text{i}2\pi \beta n m} =- \frac{\sinh x_0}{\cos 2\pi \beta n +\cosh x_0}, 
\ee
where $\cosh x_0= - \frac{E}{\Delta}$. Thus, it is found 
\be 
\Delta \left(\cos 2\pi \beta n - \frac{E}{\Delta} \right)\alpha_n=- \Delta \sinh x_0  T_n(x_0) \alpha_n.
\ee
Multiplying the above equation by $\sum_n e^{\text{i}2\pi \beta n k} $, one finally gets
\be 
\Delta \sum_n e^{\text{i}2\pi \beta n k} \left(\cos 2\pi \beta n - \frac{E}{\Delta} \right)\alpha_n = - \Delta \sinh x_0 \alpha_k.
\ee

As for the hopping item, it is noted 
\be 
\sum_{n'} \left(-1\right)^{\left|n-n'\right|} e^{-\left|n-n'\right| x } \alpha_{n'} &=& \sum_{n'}e^{-\left|n-n'\right| x } \left(-1\right)^{\left|n-n'\right|} T^{-1}_{n'}(x_0) \sum_{k'} e^{- \text{i}2\pi \beta n' k'} \alpha_{k'}   \nonumber \\
&=&  \sum_{k'} \alpha_{k'} \left(\sum_m e^{-x_0 \left| m \right|} \left(-1\right)^{\left|m \right|} e^{\text{i}2\pi \beta n' m} \right) \sum_{n'} e^{-\left|n-n'\right| x } \left(-1\right)^{\left|n-n'\right|} e^{- \text{i}2\pi \beta n' k'}\nonumber \\
&=&  \sum_{k', m }\alpha_{k'}  e^{-x_0 \left| m \right|}  \left(-1\right)^{\left|m \right|} e^{\text{i}2\pi \beta n\left( m- k'\right)} \left[\sum_{n'}e^{-\left|n-n'\right| x } \left(-1\right)^{\left|n-n'\right|}  e^{\text{i}2\pi \beta \left( m- k'\right)\left(n'-n \right)} \right] \nonumber \\
&=&   \sum_{k', m }\alpha_{k'}  e^{-x_0 \left| m \right|} \left(-1\right)^{\left|m \right|} e^{\text{i}2\pi \beta n\left( m- k'\right)}  T^{-1}_{m-k'}\left(x\right).
\ee
Multiplying the above equation by $\sum_n e^{\text{i}2\pi \beta n k} $, one obtains
\be 
&\Rightarrow & \sum_{k', m }\alpha_{k'}  e^{-x_0 \left| m \right|} \left(-1\right)^{\left|m \right|}  T^{-1}_{m-k'}\left(x\right) \sum_n e^{\text{i}2\pi \beta n\left( m- k'+k\right)} \nonumber \\
&=&   T^{-1}_{k}\left(x\right)  \sum_{k' }\alpha_{k'}  e^{-x_0 \left| k-k' \right|} \left(-1\right)^{\left|k-k' \right|},  
\ee
for which $\sum_n e^{\text{i}2\pi \beta n\left( m- k'+k\right)} = \delta\left(m- k'+k \right) $ is used.

Finally, Eq. \eqref{alpha2} can be transformed into the form
\be 
- \Delta \sinh x_0 \alpha_k -  \frac{g^2}{2J \sinh x}T^{-1}_{k}\left(x\right)  \sum_{k' }\alpha_{k'}  e^{-x_0 \left| k-k' \right|} \left(-1\right)^{\left|k-k' \right|}=0 \nonumber \\
\label{alpha2k}
\Rightarrow \Delta \sinh x_0 T_{k}\left(x\right)  \alpha_k + \frac{g^2}{2J \sinh x} \sum_{k' }\alpha_{k'}  e^{-x_0 \left| k-k' \right|} \left(-1\right)^{\left|k-k' \right|}=0. 
\ee
Evidently, Eq. \eqref{alpha2k} and Eq. \eqref{alpha2} can be transformed into each other only when  $\cosh x =\cosh x_0 $. Thus, $E_c$ satisfies 
\be 
\frac{\omega_c -E_c}{2J} =- \frac{E_c}{\Delta}.
\ee  

\subsubsection{Case C: $ \displaystyle \left|\frac{\omega_c -E }{2J}\right|<1$}
One obtains by Eq.\eqref{Hbareqn}
\be\label{alpha3} 
\Delta \left(\cos 2\pi \beta n - \frac{E}{\Delta} \right)\alpha_n - \frac{g^2}{2J \sin \theta} \sum_{n'} \sin \left(\left|n-n'\right| \theta \right) \alpha_{n'} =0.
\ee
We define $T_n\left( x_0 \right)$ as
\be
T^{-1}_n\left(\theta \right)&=& \sum_{m= -(N-1)}^{N-1}\sin \left(\left| m \right|\theta \right) e^{\text{i}2\pi \beta n m} \nonumber \\
&=& \frac{\sin N\theta \cos 2\pi\beta n \left(N-1\right) -\sin\theta - \sin\left(N-1\right) \theta \cos 2\pi\beta n N}{\cos\theta -\cos 2\pi\beta n N}.
\ee
Under $N\rightarrow \infty$, $\sin\left(N-1\right) \theta \sim \sin N \theta$ and $\cos 2\pi\beta n \left(N-1\right) \sim \cos 2\pi\beta n N$. One thus gets
\be 
T^{-1}_n\left( \theta \right)\simeq \frac{\sin\theta}{\cos 2\pi\beta n N -\cos\theta}.
\ee

Defining $\theta_0$ as $\cos\theta_0=\frac{E}{\Delta}$, one finds 
\be 
\Delta \left(\cos 2\pi \beta n - \frac{E}{\Delta} \right)\alpha_n\Rightarrow \Delta \sin\theta_0 T_n\left( \theta_0 \right) \alpha_n.
\ee
Multiplying the above equation by $\sum_n e^{\text{i}2\pi \beta n k} $, one finally gets
\be 
\Delta \sum_n e^{\text{i}2\pi \beta n k} \left(\cos 2\pi \beta n - \frac{E}{\Delta} \right)\alpha_n =\Delta \sin\theta_0 \alpha_k. 
\ee
Similarly for the hopping term in Eq. \eqref{alpha3}, it is found 
\be 
\sum_{n'} \sin \left(\left|n-n'\right| \theta \right) \alpha_{n'} &=&  \sum_{n'}\sin\left( \left|n-n'\right| \theta_0 \right) T^{-1}_{n'}(\theta_0) \sum_{k'} e^{- \text{i}2\pi \beta n' k'} \alpha_{k'}   \nonumber \\
&=& \sum_{k'} \alpha_{k'} \left(\sum_m \sin \left( \left| m \right| \theta_0\right)  e^{\text{i}2\pi \beta n' m} \right) \sum_{n'} \sin \left(\left|n-n'\right| \theta \right)  e^{- \text{i}2\pi \beta n' k'}\nonumber \\
&=&  \sum_{k', m }\alpha_{k'}  \sin \left( \left| m \right| \theta_0\right) e^{\text{i}2\pi \beta n\left( m- k'\right)} \left[\sum_{n'}\sin \left(\left|n-n'\right| \theta \right)   e^{\text{i}2\pi \beta \left( m- k'\right)\left(n'-n \right)} \right] \nonumber \\
&=&   \sum_{k', m }\alpha_{k'} \sin \left( \left| m \right| \theta_0\right) e^{\text{i}2\pi \beta n\left( m- k'\right)}  T^{-1}_{m-k'}\left(\theta\right).
\ee
Multiplying the above equation by $\sum_n e^{\text{i}2\pi \beta n k} $, one obtains
\be
 \Delta \sin \theta_0  \alpha_k - \frac{g^2}{2J \sin \theta } T^{-1}_{k}\left(\theta\right)  \sum_{k' }\alpha_{k'}  \sin \left( \left| k-k' \right| \theta_0 \right) =0 \nonumber \\
 \label{alpha3k}
\Rightarrow \Delta\sin \theta_0 T_{k}\left(\theta \right) \alpha_k-  \frac{g^2}{2J \sin \theta} \sum_{k' }\alpha_{k'}   \sin \left( \left| k-k' \right| \theta_0 \right)=0. 
\ee
It is obvious that Eq. \eqref{alpha3k} and Eq. \eqref{alpha3} can be transformed into each other when $\cos\theta=\cos\theta_0$. Thus, $E_c$ satisfies 
\be 
\frac{E_c- \omega_c}{2J} =\frac{E_c}{\Delta}.
\ee

\subsubsection{Discussion}
Interestingly, mobility edge $E_c$ for three cases satisfies the same equation in form, i.e. 
\be\label{Ec1} 
\frac{E_c- \omega_c}{2J} =\frac{E_c}{\Delta}.
\ee 
However, it should be emphasized that $E_c$ cannot be derived directly by the equation above because the relationship between $E-\omega_c$ and $2J$  imposes a fundamental restriction on  $E_c$. As shown in the previous derivations, the general duality transformations are significantly dependent on  $\left|\frac{\omega_c -E }{2J} \right| >1$ or $<1$. This feature implies that on the critical position  $\left|\frac{\omega_c -E }{2J} \right| = 1$,  all $E_c$ have to coincide each other. This requirement impose the following condition for $E_c$, i.e.
\be \label{Ec2}
\left|\frac{\omega_c -E_c }{2J} \right| = 1.
\ee

Combined Eqs. \eqref{Ec1} and \eqref{Ec2} together, the following results  can be obtained
\begin{enumerate}
\item By $\left|\frac{\omega_c -E_c }{2J} \right| = 1$, we obtain 
\be  E_c= \omega_c \pm 2J. \ee

\item By $\left| \frac{E_c}{\Delta} \right|=1 $, we obtain  
\be  E_c=\pm \Delta. \ee

\item By Eq. \eqref{Ec1}, one gets $E_c= \frac{\Delta \omega_c}{\Delta - 2J}$. Substituting the relation into   Eq. \eqref{Ec2}, we find the critical quasiperiodic potential strength  $\Delta_c$ 
    \be 
    \Delta_c= \pm \omega_c +2J.
    \ee 
\end{enumerate}
It is obvious that the former two equations actually define the four mobility edges. In contrast, the last equation gives two critical values of $\Delta$, across which the localization of the system of emitters would be altered dramatically. The validity of the results  is examined  by evaluating the localization of effective eigenstate $\overline{\alpha}$. As shown in Fig. \ref{fig:IPRandFD},  the $E-\Delta$ plane can be divided into several regions by the relationships above, which respectively characterize the distinct localizations for the system of emitters. 

\begin{figure}
\center
\includegraphics[height=17cm]{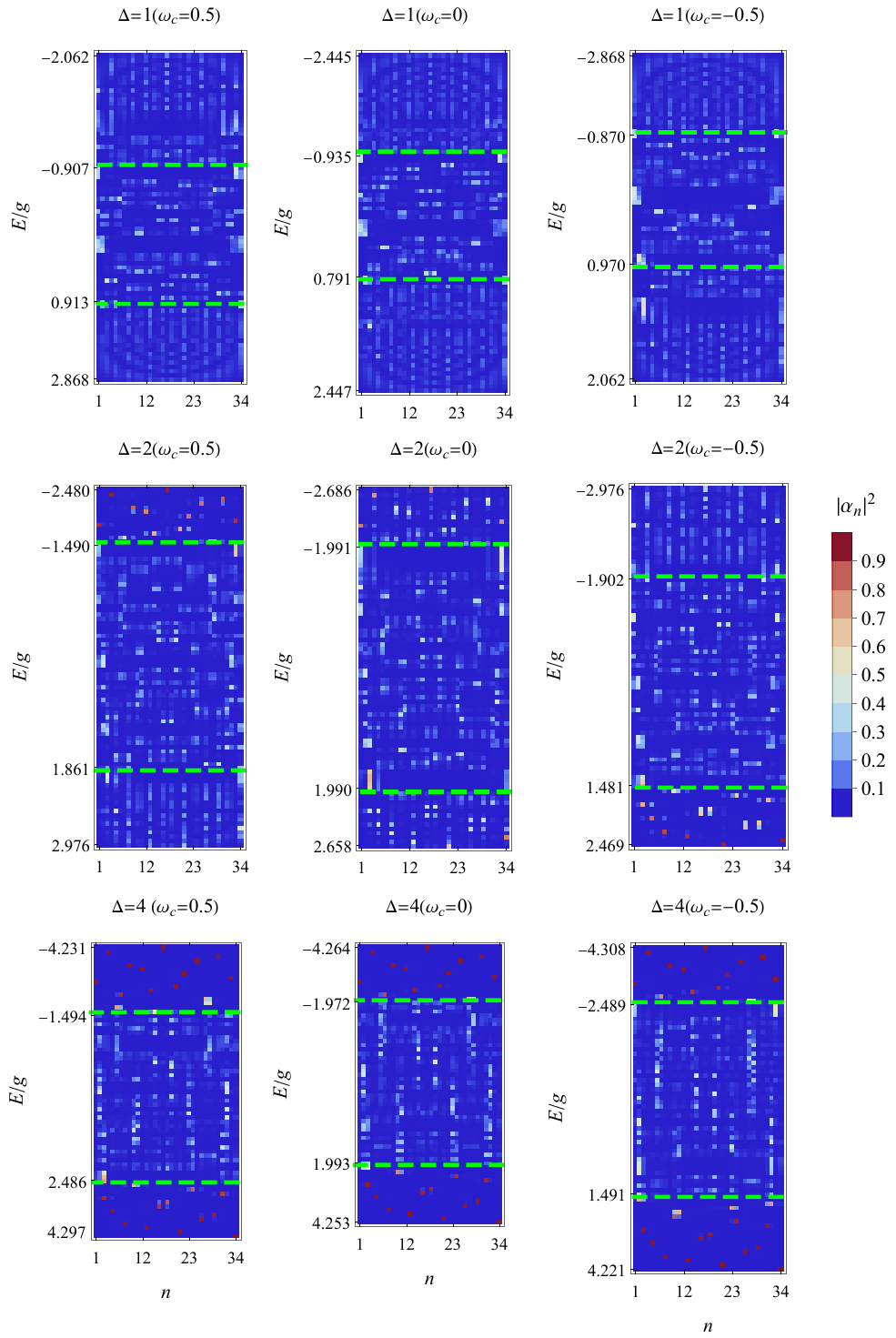}
\caption{ The population $\left| \alpha_n \right|^2$ of  bound states for $\omega_c=0.5, 0, -0.5$ when $\Delta=1, 2, 4$. The dashed-green lines depict the values of $E$ closed to the edge $E_c$. For all plots, $N=34$ and $g=1$ are chosen. }
\label{figS:boundstate}
\end{figure}

Additionally, in Fig. \ref{figS:boundstate},  the population $\left| \alpha_n \right|^2$  are plotted for the bound states when  $\omega_c=0.5, 0, -0.5$ and  $\Delta=1, 2, 4$ respectively. To illustrate the influence of ME, the level $E$ closed to $E_c$ is highlighted by dashed-green lines.  It is evident that the bound states display distinct localization properties  across $E_c$. For an instance of $\Delta=1$, the bound states of $\left| E \right|>1$ are manifestly extended since the value of  $\left| \alpha_n \right|^2$ tends to be isotropic for all sites. In contrast for $\left| E \right|<1$,  $\left| \alpha_n \right|^2$ can be pronounced on a few isolated sites, an typical  characterization for critical states \cite{zhou26}. The central frequency $\omega_c$ can also alter the localization of  bound states. As shown by the middle panels with $\Delta=2$, the extended  or localized  states can occur at the high or low levels, decided by  $\omega_c=0.5$ or $-0.5$. This picture can be attributed to the variance of onsite potentials, induced by $\overline{H}_{\text{int}}$. As illustrated by Eq. \eqref{effectHs} for $\left|\frac{\omega_c -E }{2J}\right|>1$,  the potential can be altered by amount of  $\pm 1/\sinh x$.

\subsection{The effect of finite $N$}

\begin{figure}
\center
\includegraphics[width=16cm]{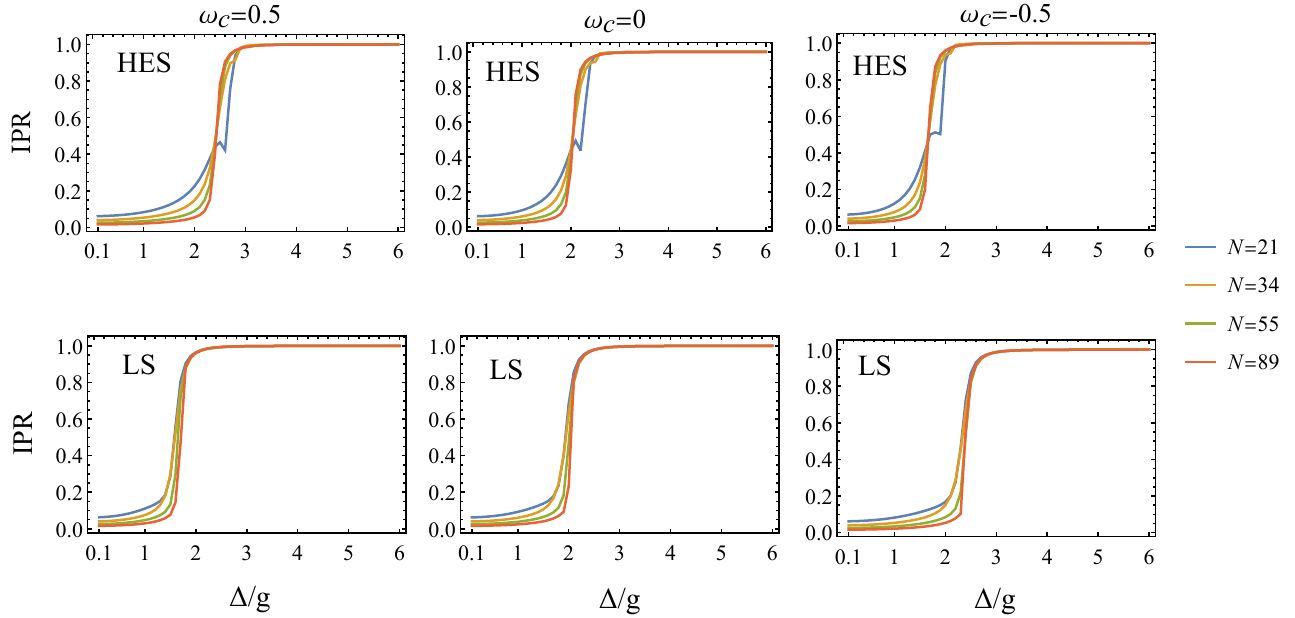}
\caption{ The IPR of HES or LS for $N=21, 34, 55, 89$ when $\omega_c=0.5, 0, -0.5$ respectively.  For all plots, $g=1$ is chosen. }
\label{figS:scaling}
\end{figure}

It is noted for Fig. \ref{fig:IPRandFD}  that both IPR and $\tau$ seemingly display strong fluctuation near $E_c$ and critical $\Delta$.  With respect that $E_c$ and critical $\Delta$ is determined under $N\rightarrow \infty$, this deficiency comes from  the effect of finite $N$. In Fig. \ref{figS:scaling}, IPRs for the highest excited state (HES) and lowest state (LS) are plotted for $N=21, 34, 55, 89$ respectively. Evidently, the variance of IPR  across $\Delta_c$ becomes much steep with the increasing of $N$. Thus, it is expected that the deficiency can be disappear for a large $N$. It is admitted that the evaluation would be exhaustive when $N$ is large. Especially for the determination of BICs occurring in the region $\left|\frac{\omega_c -E }{2J}\right|<1$, the distribution of solution $E$s becomes so dense that we have to shorten significantly the step length of searching to find all solutions, which make the computation exhaustive.    

\subsection{Occurrence of the edge states around $E=0$ when $\Delta=0$ }\label{SM:edgestate} 
\begin{figure}
\center
\includegraphics[width=18cm]{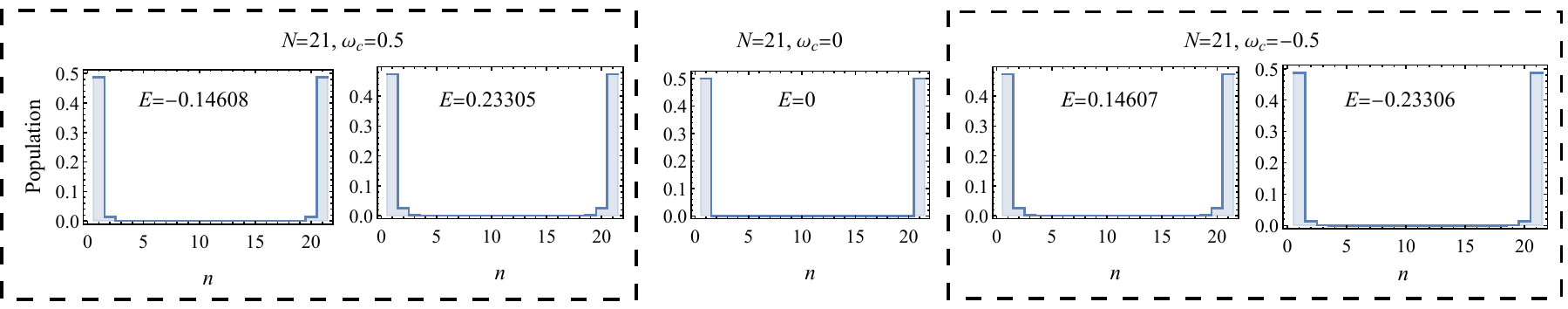}
\includegraphics[width=16cm]{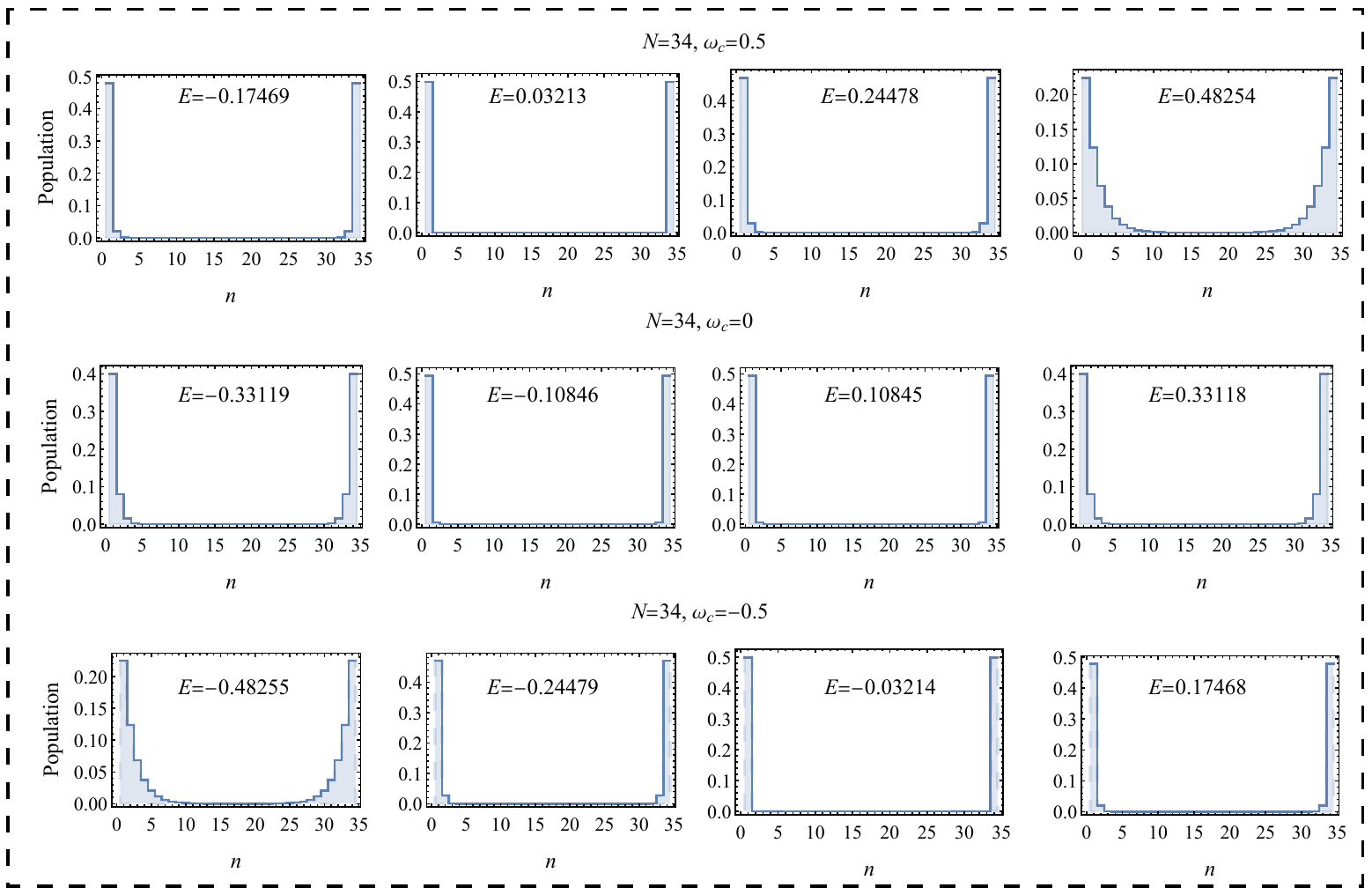}
\caption{ The population $\left|\alpha_n \right|^2$ of the bound state around $E=0$ for $\omega_c=0.5, 0, -0.5$ when $\Delta=0$. For the plots, $N=21$ and $34$ are chosen respectively, for which the edge state $\ket{E=0}$ can occur for the former case.  $g=1$ is  adopted for all plots and the labelled energy $E$  is scaled in unit of $g$.  }
\label{figS:edgestate}
\end{figure}

The explicit calculation shows that the edge state can occur exactly at $E=0$ under open boundary condition when $\Delta=\omega_c=0$ with odd $N$. In this circumstance,  $\overline{H}_{s}$ becomes
\be 
\overline{H}_{s}&=&- \frac{g^2}{2J} \sum_{n, n'} \sigma^{+}_n \sigma^{-}_{n'}  \sin \frac{\left|n-n'\right|  \pi}{2}\nonumber \\
&=&- \frac{g^2}{2J} \sum_{n=1}^{N} \sum_{\text{odd} \delta n=1}^{N-1}\left(-1 \right)^{\left(\delta n -1\right)/2} \left( \sigma^{+}_n \sigma^{-}_{n+ \delta n} + \sigma^{+}_n \sigma^{-}_{n-\delta n} + h. c. \right). 
\ee
It is obvious that  the hopping can not only happen uniquely between the emitters with odd-site distance, but also is manipulated by a phase factor $\left(-1 \right)^{\left(\delta n -1\right)/2}$. Consequently, for the excitation on the boundaries when $N=2\overline{N}+1$ ($\overline{N}$ being integer number), one gets
\be 
\overline{H}_{s} \sigma^{\dagger}_1 \ket{0}= - \frac{g^2}{2J} \sum_{n=1}^{\overline{N}} \left( -1 \right)^{n+1} \sigma^{\dagger}_{2n}\ket{0}, \overline{H}_{s} \sigma^{\dagger}_N \ket{0}= -\left( -1 \right)^{\overline{N}} \frac{g^2}{2J} \sum_{n=1}^{\overline{N}} \left( -1 \right)^{n} \sigma^{\dagger}_{2n}\ket{0}.
\ee
By the relations above, it is direct to find 
\be 
\overline{H}_{s} \left[ \sigma^{\dagger}_1 + \left( -1 \right)^{\overline{N}} \sigma^{\dagger}_N \right] \ket{0}=0.
\ee
Thus, the edge state is defined as 
\be 
\ket{E=0}= \frac{1}{\sqrt{2}} \left[ \sigma^{\dagger}_1 + \left( -1 \right)^{\overline{N}} \sigma^{\dagger}_N \right] \ket{0},
\ee
for which the excitation becomes located entirely at the  end sites with equal probability. 

It should be emphasized that the nonvanishing $\Delta$ or $\omega_c$ may lead to the deviation of bound state away from $E=0$. As a result, the population of excitation behaves exponentially extended, as shown in Fig. \ref{figS:edgestate}.    
   

\begin{thebibliography}{99}
\bibitem{abanin}D. A. Abanin, E. Altman, I. Bloch, and M. Serbyn, Colloquium: Many-body localization, thermlization, and entanglment, Rev. Mod. Phys. {\bf 91}, 021001 (2019).

\bibitem{aa} S. Aubry and G. Andr\'{e},  Analyticity breaking and Anderson localization in incommensurate lattices, Ann. Isr. Phys. Soc. {\bf 3}, 133 (1980).

\bibitem{haper} P. G. Haper, Single band motion of conduction electrons in a uniform magnetic field, Proc. Phys. Soc. London Sect. A {\bf 68}, 874 (1955)
    
\bibitem{basko}D. M. Basko, I. L. Aleeiner, and B. L. Altshuler, Metal-insulator transition in  a weakly interacting many-electron system with localized single-particle states, Ann. Phys. (Amsterdam) {\bf 321}, 1126-1205 (2006). 
    
\bibitem{huse} V. Oganesyan and David A. Huse, Localization of interaction fermions at high temperature, Phys. Rev. B {\bf 75}, 155111 (2007). 
    
\bibitem{sarma90}S. Das Sarma, S. He, and X. C. Xie, Localization, mobility edge, and metal-insulator transition in a class of one-dimensional slowly varying deterministic potentials, Phys. Rev. B {\bf 41}, 5544-5565 (1990).
    
\bibitem{biddle}J. Biddle and S. Das Sarma, Predicted Mobility edge in one-dimensional incommensurate optical lattices: an exactly solvable model of anderson localization, Phys. Rev. Lett. {\bf 104}, 070601 (2010).
    
\bibitem{ganeshan} S. Ganeshan, J. H. Pixley, and S. Das Sarma, Nearest neighbor tight binding models with an exact mobility edge in one dimension, Phys. Rev. Lett. {\bf 114}, 146601 (2015).   
    
\bibitem{wang20} Yucheng Wang, Xu Xiu, Long Zhang, Hpeng Yao, Shu Chen, Jiangong You, Qi Zhou, and Xiong-Jun Liu, One-dimensional quasiperiodic mosaic lattice with exact mobility edge, Phys. Rev. Lett. {\bf 125}, 196604 (2020). 

\bibitem{wang16}Jun Wang, Xia-Ji Liu, Gao Xianlong, and Hui Hu, Phase diagram of a non-Abelian Aubry-Andr\'{e}-Harper model with p-wave superfluidity, Phys. Rev. B 93, 104504 (2016).

\bibitem{wang21}Y. Wang, C. Cheng, X.-J. Liu, and D. Yu, Many-body critical phase: extended and nonthermal, Phys. Rev. Lett. {\bf 126}, 080602 (2021).

\bibitem{wang22}Y. Wang, L. Zhang, W. Sun, T.-F. J. Poon, and X.-J. Liu, Quantum phase with coexisting localized, extended, and critical zones, Phys. Rev. B {\bf 106}, L140203 (2022).
    
\bibitem{liu22}T. Liu, X. Xia, S. Longhi, and L. Sanchez-Palencia, Anomalous mobility edges in one-dimensional quasiperiodic models, SciPost Phys. {\bf 12}, 027 (2022).
    
\bibitem{zhou23}X.-C. Zhou, Y. Wang, T.-F. J. Poon, Q. Zhou, and X.-J. Liu, Exact new mobility edges between critical and localized states, Phys. Rev. Lett. {\bf 131}, 176401 (2023).
    
\bibitem{goncalves23}M. Gon\c{c}alves, B. Amorim, E. V. Castro, and P. Ribeiro, Critical Phase Dualities in 1D Exactly Solvable Quasiperiodic Models, Phys. Rev. Lett. {\bf 131}, 186303 (2023).
    
\bibitem{li23}Hao Li, Yong-Yi Wang, Yun-Hao Shi, Kaixuan Huang, Xiaohui Song, Gui-Han Liang, Zheng-Yang Mei, Bozhen Zhou, He Zhang, Jia-Chi Zhang, Shu Chen, S. P. Zhao, Ye Tian, Zhan-Ying Yang, Zhongcheng Xiang, Kai Xu, Dongning Zheng and Heng Fan, Observation of critical phase transition in a generalized Aubry-André-Harper model with superconducting circuits, npj Quantum Information, {\bf 9}, 40 (2023).

\bibitem{banerjee} S. Banerjee, S. R. Padhi, and T. Mishra, Emergence of distinct exact mobility edges in a quasiperiodic chain, Phys. Rev. B {\bf 111}, L220201 (2025).
    
\bibitem{zhou26}Xin-Chi Zhou, Bing-Chen Wang, Yongjian Wang, Yucheng Wang, Yudong Wei, Qi Zhou, and Xiong-Jun Liu, The fundalmental localization phases in quasiperiodic systems: a unified framework and exact results, Science Bulletin in press (2026).
    
\bibitem{huang26} Wenhui Huang, Xin-Chi Zhou, Libo Zhang, Jiawei Zhang, Yuxuan Zhou, Bing-Chen Yao, Zechen Guo, Peisheng Huang, Qixian Li, Yongqi Liang, Yiting Liu, Jiawei Qiu, Daxiong Sun, Xuandong Sun, Zilin Wang, Changrong Xie, Yuzhe Xiong, Xiaohan Yang, Jiajian Zhang, Zihao Zhang, Ji Chu, Weijie Guo, Ji Jiang, Xiayu Linpeng, Wenhui Ren, Yuefeng Yuan, Jingjing Niu, Ziyu Tao, Song Liu, Youpeng Zhong, Xiong-Jun Liu, Dapeng Yu, Experimental observation of exact quantum critical states, Nature Physics in press (2026).

\bibitem{hu26}Zhongshu Hu, Yajing Guo, Yu-Dong Wei, Bing-Chen Yao, Zhentian Qian, Xin-Chi Zhou, Bao-Zong Wang, Jianing Yang, Xuzong Chen, Shengjie Jin, Xiong-Jun Liu, Observation of a tripartite quantum phase for coexisting extended, localized, and critical states, arXiv: 2605.21441 (2026).
    
\bibitem{liu24}Yaru Liu, Zeqing Wang, Chao Yang, Jianwen Jie, and Yucheng Wang, Dissipation induced extended-localized transition, Phys. Rev. Lett. {\bf 132}, 216301 (2024). 
    
\bibitem{longhi}S. Longhi, Dephasing-induced mobility edges in a quasicrystals, Phys. Rev. Lett. {\bf 132}, 236301 (2024). 

\bibitem{greentree06}A. D. Greentree, C. Tahan, J. H. Cole, and C. L. Hollenberg, Quantum phase transitions of light, Nat. Phys.{\bf 2}, 856–861 (2006).
    
 \bibitem{angelakis07}D. G. Angelakis, M. F. Santos, and S. Bose, Photonblockade-induced Mott transitions and XY spin models in coupled cavity arrays, Phys. Rev. A {\bf 76}, 031805(R) (2007).    

\bibitem{john91}J. Wang and S. John, Quantum optics of localized light in a photonic band gap, Phys. Rev. B {\bf 43}, 12772 (1991).

\bibitem{kofman94}A. G. Kofman, G. Kurizki, and B. Sherman, Spontaneous and induced atomic decay in phontic band structures, J. Mod. Opt. {\bf 41}, 353 (1994). 
    
\bibitem{calojo} G. Gcalj\'{o}, F. Ciccarella, D. Chang, and P. Rabl, Atom-field dressed states in a slow-light wavegude QED, Phys. Rev. A {\bf 93}, 033833 (2016).

\bibitem{marinica}D. C. Marinica, A. G. Borisov, and S. V. Shabanov, Bound state in the continuum in photonics, Phys. Rev. Lett. {\bf 100}, 183902 (2008).
    
\bibitem{bulgakov}Evgeny N. Bulgakov and Almas F. Sadreev, Bound states in the continuum in photonic waveguides inspired by defects,  Phys. Rev. B {\bf 78}, 075105 (2008).
    
\bibitem{plotnik}Y. Plotnik, Or Peleg, F. Dreisow, M. Heinrich, S. Nolte, A. Szameit, and M. Segev, Experimental Observation of Optical Bound States in the Continuum, Phys. Rev. Lett. {\bf 107}, 183901 (2011).
    
\bibitem{hu25a}Hai-Tao Hu, Xiaoshui Lin, Ai-Min Guo, Guangcan Guo, Zijin Lin, Ming Gong, Hidden self-duality and exact mobility edges in quasiperiodic network models, 	Phys. Rev. Lett. {\bf 134}, 246301 (2025).
    
\bibitem{hu25b}Hai-Tao Hu, Yang Chen, Xiaoshui Lin, Ai-Min Guo, Zijing Lin, Ming Gong, Exact mobility edges in quasiperiodic network models with slowly varying potentials, Phys. Rev. B {\bf 112}, 054201 (2025).
    
\bibitem{cui25} H. T. Cui, Y. A. Yan, M. Qin, and X. X. Yi, Effective Hamiltonian approach to the exact dynamics of open system by complex discretization approximation for environment, APL Quantum, {\bf 2}, 026116 (2025). 
    
\bibitem{wang25} Jian Wang, Dong-Yu Huang, Xiao-Long Zhou, Ze-Min Shen, Si-Jian He1, Qi-Yang Huang, Yi-Jia Liu, Chuan-Feng Li, and Guang-Can Guo, Ultrafast High-Fidelity State Readout of Single Neutral Atom, Phys. Rev. Lett. {\bf 134}, 240802 (2025).
    
\bibitem{shaw26}Adam L. Shaw, Anna Soper, Danial Shadmany, Aishwarya Kumar, Lukas Palm, Da-Yeon Koh, Vassilios Kaxiras, Lavanya Taneja, Matt Jaffe, David I. Schuster and  Jonathan Simon, A cavity-array microscope for parallel single-atom interfacing, Nature {\bf  650}, 320–326 (2026).

\bibitem{imamoglu}A. Imamo\={g}lu, H. Schmidt, G. Woods,  and M. Deutsch, Strongly Interacting Photons in a Nonlinear Cavity, Phys. Rev. Lett. {\bf 79}, 1467-1470 (1997).
    
\bibitem{modak15} R. Modak and S. Mukerjee, Many-body localization in the presence of single-partical mobility edge, Phys. Rev. Lett. {\bf 115}, 230401 (2015).
    
\bibitem{nag17}S. Nag and A. Garg, Many-body mobility edges in a one-dimensional system of interaction fermions, Phys. Rev. B {\bf 96}, 060203(R) (2017).
    
\bibitem{alexan18} F. Alex An, Eric J. Meier, and B. Gadway, Engineering a flux-dependent mobility edge in disordered zigzag chains, Phys. Rev. X {\bf 8}, 031045 (2018).
    
\bibitem{kohlert}T. Kohlert, S. Scherg, X. Li, Henrik P. L\"{u}schen, S. D. Sarma, I. Bloch, and M. Aidelsburger, Observation of many-body localization in a one-dimensional system with a single-particle mobility edge, Phys. Rev. Lett. {\bf 122}, 170403 (2019).
    
\bibitem{wei19} Xingbo Wei, Chen Cheng, Gao Xianlong, and R. Mondaini,  Investigating many-body mobility edges in isoloated quantum systems, Phys. Rev. B {\bf 99}, 165137 (2019).
    
\bibitem{alexan21} F. Alex An, K. Padavi\'{c}, Eric J. Meier, S. Hegde, S. Ganeshan, J. H. Pixley, S. Vishveshwara, and B. Gadway, Interaction and mobility edges: observing the generlized Aubry-Andr\'{e} model, Phys. Rev. Lett. {\bf 126}, 040603 (2021).
    
\bibitem{wangyunfei21} Y. Wang, J.-H. Zhang, Y. Li, J. Wu, W. Liu, F. Mei, Y. Hu, L. Xiao, J. Ma, C. Chin, and S. Jia, Observation of interaction-induced mobility edge in a disordered atomic wire, Phys. Rev. Lett. {\bf 129}, 130401 (2021).

\bibitem{huang23} K. Huang, D. Vu, X. Li, and S. Das Sarma, Incommensurate many-body localization in the presence of long-range hopping and single-particle mobility edge, Phys. Rev. B {\bf 107}, 035129 (2023).
    
\bibitem{huang24}K Huang, D. Vu, S. Das Sarma, and X. Li, Interaction-enhanced many-body localization in a generalized Aubry-Andr\'{e} model, Phys. Rev. Research {\bf 6}, L022054 (2024).
    
\bibitem{winkler06}K. Winkler, G. Thalhammer, F. Lang, R. Grimm, J. Hecker Denschlag, A. J. Daley, A. Kantian, H. P. B\"{u}chler, and P. Zoller, Repulsively bound atom pairs in an optical lattice, Nature, {\bf  441}, 853–856 (2006) 
\end{thebibliography}
\end{document}